\documentstyle[prl,aps,eqsecnum,epsf]{revtex}
\input{epsf}
\begin{document}
\title{A Measure of Strength of an Unextendible Product Basis } 
\author{S. Chaturvedi\thanks{email: scsp@uohyd.ernet.in}}
\address{ Department of Physics, University of Hyderabad, Hyderabad 500046,
India}
\date{\today}
\maketitle
\begin{abstract}
A notion of strength of an unextendible product basis is introduced
and a quantitative measure for it is 
suggested with a view to providing an indirect measure for the bound 
entanglement of formation of the bound entangled mixed state 
associated with an unextendible product basis.
\end{abstract}
\vspace{0.3cm}
Quantum nonlocality\cite{einstein} has, for long,  owing to well known
historical reasons, been associated  with entangled states - states in a
tensor product Hilbert space with which are not of a product type. Such
states, which have not only generated numerous fascinating and intricate
mathematical concepts and questions pertaining to identification,
quantification and classification of entanglement, but have also
played a key role in the development of such novel ideas as 
quantum teleportation\cite{bennett1}, quantum
cryptography\cite{bennett2}, quantum dense coding \cite{bennett3}
and quantum computation \cite{nielsen} some of which
have aso been experimentally realised\cite{expts}.
Recent years have seen the advent of a new kind of
nonlocality- nonlocality without entanglement\cite{bennett4},
associated with, not just a single state, but with a set of orthogonal
product states on a tensor product Hilbert space such that there is no
state of a product type orthogonal to all the members of the set.
The nonlocality associated with such a set, referred to as an unextendible
product basis(UPB)\cite{bennett5}, manifests itself through the impossibility of distinguishing between
the members of the set through local operations and classical communication.
In contrast to the nonlocality associated with the entangled states, which
arises at the level of states in tensor product Hilbert spaces, that
associated with a UPB may be thought of as a reflection of the non
commutativity at the level of operators
\cite{diVincenzo2}. Besides exhibiting nonlocality without entanglement,
UPB's have ramifications for entanglement as well, in that they enable
explicit constructions of bound entangled mixed states\cite{horodecki}-
entangled mixed states with a positive Peres transpose\cite{peres}.
The very first explicit examples of UPB's, that appeared in the pioneering
work of Bennett et al\cite{bennett5}, were -
the {\bf Pyramid} and the
{\bf Tiles}. These set of states constitute minimal UPB's (UPB's with the
smallest
permissible dimension) on ${\cal H}_3 \bigotimes {\cal H}_3$ and are
explicitly given below:
\begin{itemize}
\item{\bf Pyramid}:
$\psi_i = v_i\otimes w_i~~;~~i=0,\dots,4 $~
where
\begin{eqnarray}
v_i &=& N(\cos\frac{2\pi i}{5},\sin\frac{2\pi i}{5},h) ~~~;~~~
h=\sqrt{-\cos\frac{4\pi} {5}}~~; ~~ N=\frac{1}{\sqrt{1+|\cos\frac{4\pi}{5}|}},
\\
w_i &=&  v_{2i~mod~5}~
\end{eqnarray}
\item{\bf Tiles}
$\psi_i = v_i \otimes w_i~~~;~~~i=0,\cdots,4$ 
~where
\begin{eqnarray}
& &v_0=|0>;v_1= \frac{1}{\sqrt{2}}(|0>-|1>);v_2=|2>;v_3= 
\frac{1}{\sqrt{2}}(|1>-|2>);v_4=\frac{1}{\sqrt{3}}(|0>+|1>+|2>), \\
& &w_0= \frac{1}{\sqrt{2}}(|0>-|1>);w_1=|2>; 
w_2=\frac{1}{\sqrt{2}}(|1>-|2>);w_3= |0>; 
w_4=\frac{1}{\sqrt{3}}(|0>+|1>+|2>).
\end{eqnarray}
\end{itemize}
In a subsequent work, DiVincenzo et al \cite{diVincenzo1} gave
general constructions of several UPB's which are listed below 
\begin{itemize}
\item {\bf Gen Shifts}:$2k$-dimensional (minimal) UPB on
$\bigotimes_{i=1}^{2k-1}{\cal H}_2$.
\item {\bf Gen Pyramids}:$p$-dimensional (minimal) UPB on
$\bigotimes_{i=1}^{n}{\cal H}_3 $ with $n$ such that $2n+1=p$, a prime.
\item{\bf Quad Res}: $p$-dimensional (minimal) UPB on
${\cal H}_n\bigotimes{\cal H}_n $ with $n$ such that $2n-1=p$, a prime of
the form $4m+1$. 
\item{\bf Gen Tiles1}:$(n-1)^2$-dimensional(non minimal) UPB on
${\cal H}_n\bigotimes{\cal H}_n $ with $n$ even. 
\item {\bf Gen Tiles2}: $(nm-2m+1)$- dimensional (non minimal) UPB on
${\cal H}_m\bigotimes{\cal H}_n$ with $n\geq m~;~n>3~;~ m\geq 3$.
\end{itemize} 
The present work is inspired by a result quoted in\cite{bennett5} that
while the bound entangled of formation\cite{horodecki} for the
mixed state associated with the {\bf Pyramid} UPB is 0.232635 ebits, that
for the state associated with the Tiles UPB turns out to be 0.213726 ebits.
In view of the fact that the calculation of these numbers involves
extensive numerical searches, one is led to the question whether it is
possible to associate with each UPB a number as a kind of measure of
strength of that UPB which would, in turn, provide an indirect measure of
the bound entanglement of formation of the associated bound entangled mixed
state. This work is an effort in this direction.     

To motivate the definition of the strength of a UPB, we begin by examining
the structure of the scalar products among the $v_i$'s and the $w_i$'s for
the {\bf Pyramid}. This UPB is characterized by  $(v_0,v_2)=(v_2,v_4)=
(v_4,v_1)=(v_1,v_3)=(v_3,v_0) = 0;(w_0,w_1) =(w_1,w_2)=(w_2,w_3)=(w_3,w_4)
=(w_4,w_0)=0$ with the remaining scalar products non zero. If any of the nonzero
scalar products were to vanish, the resulting set will not be a UPB.
It seems, therefore natural to define the strength $s$ of a UPB 
by the the magnitude of the product of the non-zero scalar products among
the $v_i$'s and the $w_i$'s. In the particular case of the {\bf Pyramid},
the resulting expression can be compactly written as
$s= |B_5(v_0,v_1,v_2,v_3,v_4)\times B_5(w_0,w_2,w_4,w_1,w_3)|$.
Here $B_5(v_0,v_1,v_2,v_3,v_4)\equiv (v_0,v_1)(v_1,v_2)(v_2,v_3)(v_3,v_4)
(v_4,v_0)$ and $B_5(w_0,w_2,w_4,w_1,w_3)$ is similarly defined.
(These objects  can be identified with the fifth order Bargmann
invariants \cite{bargmann} associated with the set of vectors
$v_i, w_i, i=0,\cdots 4$ repectively) The value of $s$ 
turns out to be  $[(30\sqrt{5}-66)/12]^2$.

Similarly, for the {\bf Tiles}, we have
$(v_0,v_3)=(v_3,v_4)=(v_4,v_1)=(v_1,v_2)=(v_2,v_0) = 0;
(w_0,w_1)=(w_1,w_3)=(w_3,w_2)=(w_2,w_4)=(w_4,w_0) = 0$ with the remaining
scalar products non zero. The strength $s$ of this UPB can be written as
 $s=|B_5(v_0,v_1,v_3,v_2,v_4)]\times B_5(w_0,w_2,w_1,w_4,w_3)|$  
and its value turns out to be $(1/12)^2$, which is less than that for the
{\bf Pyramid}. Thus we find that the {\bf Pyramid}, in this sense, is
stronger than the {\bf Tiles} and one is tempted to conclude that it is
this strength which leads to a higher value for the bound entanglement of
formation of the associated bound entangled mixed state vis a vis the
{\bf Tiles}.

To probe further, the connection suggested above, between the strength and the
entropy of bound entanglement of formation of the associated bound
entangled mixed state, we examine the six parameter family of UPB on
${\cal H}_3 \bigotimes {\cal H}_3$ constructed by DiVincenzo et al
\cite{diVincenzo1}
\begin{eqnarray}
v_0 &=& |1>,\nonumber\\
v_1 &=& \sin\gamma_B \sin\theta_B |0> - \sin\gamma_B \cos\theta_B |2> 
+\cos\gamma_B e^{i\phi_B} |1>,\nonumber\\ 
v_2 &=& |0>,\\
v_3 &=& \cos\theta_B |0> +\sin\theta_B |2>,\nonumber\\ 
v_4 &=& \frac{1}{N_B}(\sin\gamma_B \cos\theta_B e^{i\phi_B}|1> +
\cos\gamma_B |2>. \nonumber\\
w_0 & =& |0>,\nonumber\\
w_1 &=& |1>,\nonumber\\
w_2 &=& \cos\theta_A |0> +\sin\theta_A |2>, \\ 
w_3 &=& \sin\gamma_A \sin\theta_A |0> +\cos\gamma_A e^{i\phi_A} |1> 
-\sin\gamma_A\cos\theta_A |2>,\nonumber\\
w_4 &=& \frac{1}{N_A}(\sin\gamma_A \cos\theta_A e^{i\phi_A}|1> +  
+\cos\gamma_A |2>.\nonumber 
\end{eqnarray}
where $N_A,B = \sqrt{\cos^2 \gamma_{A,B} +\sin^2 \gamma_{A,B}\cos^2
\theta_{A,B}}$. For this to be a UPB, we must have $\cos\theta_{A,B}\neq 0, 
\cos\gamma_{A,B}\neq 0,\sin\theta_{A,B}\neq 0$. Further, as noted by 
DiVincenzo et al\cite{diVincenzo1}, this family contains {\bf Pyramid}
and {\bf Tiles} UPB's as special cases corresponding to
$\phi_{A,B}=0, \theta_{A,B}= \gamma_{A,B} = \cos^{-1}((\sqrt{5}-1)/2)$
and $\phi_{A,B}=0, \theta_{A,B}=\gamma_{A,B} =3\pi/4$.
We now ask the question whether the {\bf Pyramid}, is in some sense, a
privilleged member of this family. To this end, we
compute the expression for the strength of this family of UPB's and 
find that
\begin{eqnarray}
s&=&|B_5(v_0,v_1,v_3,v_2,v_4)]\times B_5(w_0,w_2,w_1,w_4,w_3)] \nonumber\\ 
&=& \left[\frac{\sin^2\theta_A \sin^2\gamma_A\cos^2\theta_A\cos^2\gamma_A}
{\cos^2\gamma_A+\sin^2\gamma_A\cos^2\theta_A}\right]
\times\left[\frac{\sin^2\theta_B \sin^2\gamma_B\cos^2\theta_B\cos^2\gamma_B}
{\cos^2\gamma_B+\sin^2\gamma_B\cos^2\theta_B}\right]
\end{eqnarray}
Note that $s$ is independent of the phases $\phi_{A,B}$. Setting all angles 
equal, $\theta_{A,B}=\gamma_{A,B} =\theta$ we find that the resulting 
expression has a maximum precisely at  
$x\equiv\cos\theta=(\sqrt{5}-1)/2$ which, as noted above,
corresponds to the {\bf Pyramid}. Thus, within the family of the UPB's
considered, the {\bf Pyramid} has a unique position in that it has the
maximum strength and, if the connection to the bound entanglement of
formation suggested above is correct, then one should find that the
bound entangled mixed state corresponding to the {\bf Pyramid} has the
highest entropy of bound entanglement of formation within this family.

Next, we turn to UPB's on ${\cal H}_3 \bigotimes {\cal H}_3 \bigotimes 
{\cal H}_3$. The UPB's constructed by DiVincenzo et al\cite{diVincenzo1}
are:
\begin{eqnarray}
\psi_i &=& v_i\otimes v_{2i~mod~7}\otimes v_{3i~mod~7}~~;~~i=0,\cdots,6 \\     
v_i &=& N_7(\cos\frac{2\pi i}{7},\sin\frac{2\pi i}{7},h_7) ~~~;~~~
h_7=\sqrt{-\cos\frac{2m\pi} {7}}~~; ~~ N_7=
\frac{1}{\sqrt{1+|\cos\frac{2m\pi}{7}|}}.
\end{eqnarray}
Here $m$ takes two values $2$ and $3$ and thus we have two UPB's. The
UPB corresponding to $m=2$ is referred to as the {\bf Sept}.
We now wish to locate their places within the family of UPB's on
${\cal H}_3 \bigotimes {\cal H}_3 \bigotimes {\cal H}_3$ on the basis
of their strengths. To this end, we have
constructed a parametrized family of UPB's on
${\cal H}_3 \bigotimes {\cal H}_3 \bigotimes 
{\cal H}_3$. Its members are as follows: 
\begin{eqnarray}
\psi_i &=& v_i \otimes w_i \otimes u_i ~~~;~~ i=0,\cdots, 6 \\
v_0 &=&|0>,\nonumber\\
v_1&=& N[(\sin\theta_4\cos\theta_2\cos\theta_3 e^{i(\lambda- \chi)}
-\sin\theta_4\cos\theta_1\sin\theta_2\sin\theta_3)|0>\nonumber\\
&+&(\sin\theta_3\cos\theta_1\cos\theta_4 e^{-i\mu} -\sin\theta_3
\sin\theta_1\cos\theta_2\sin\theta_4 e^{-i\chi})|1>
+(\cos\theta_3\cos\theta_4 e^{i(\lambda-\mu)}
-\sin\theta_1\sin\theta_2\sin\theta_3\sin\theta_4)|2>],\nonumber\\
v_2&=&|1>,\nonumber\\
v_3&=&\cos\theta_4 e^{i\mu}|0> +\sin\theta_4\sin\theta_2|1>-
\sin\theta_4\cos\theta_2 e^{i\chi}|2>,\\
v_4&=& \cos\theta_1|0> + \sin\theta_1 |2>,\nonumber\\
v_5&=& \cos\theta_2 |1>+ \sin\theta_2 e^{i\chi}|2>,\nonumber\\
v_6&=& \sin\theta_3\sin\theta_1|0> + \cos\theta_3 e^{-i\lambda} |1> -
\sin\theta_3\cos\theta_1 |2>.\nonumber
\end{eqnarray}
The vectors $w_i$ and $u_i$ are $w_i=v_{2i~mod~7}$ and $u_i=v_{3i~mod~7}$
respectively with a different set of parameters in each case making a total  
of $21$ parameters specifying the family. This is obviously a rather large
family. To keep matters simple, we consider a sub-family wherein the
$u$'s and $w$'s have the same parameters as those in the $v$'s. For this
sub-family it turns out the strength $s=|B_7(v_0,v_1,v_2,v_3,v_4,v_5,v_6)]
\times[B_7(v_0,v_3,v_6,v_2,v_5,v_1,v_4)]|^3$ depends on the phases
$\lambda, \mu$ and $\chi$ only through $\alpha=\lambda-\chi$ and
$\beta=\mu-\chi$. We now set all the angles equal to $\theta$ and
$\alpha=\beta$ and obtain, for $s$, the following expression
\begin{equation}
s=[f(x,y)]^3
\end{equation}
where
\begin{equation}
f(x,y)= |[\frac{[x^9 (1-x^2)^6\sqrt{4+4xy+x^2}(x^4-x^2+1+2xy(x^2-1))^2
(x^6+4x^4-4x^2+1+2x^3y(2x^2-1)]}{[(x^4-3x^2+1)^2+2x^2(1-x^2)
(x^4-x^2+1+2xy(x^2-1)]^2}]|
\end{equation}
with $x=\cos\theta$ and $y=\cos\alpha$. The function $f(x,y)$ is plotted in
Fig 1 for $-1\leq x \leq 1$ and $0 \leq y \leq 1$. (We have restricted the
range of $y$ to $0 \leq y \leq 1$ owing to the symmetry $f(x,y),=f(-x,-y)$)
\begin{figure}[h]
\begin{center}
\leavevmode
\epsfxsize=3.0in
\epsfbox{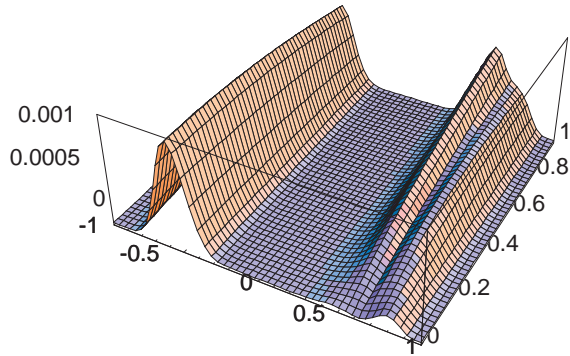}
\end{center}
\caption{Strength as a function of $x$ and $1-y$}
\end{figure}

The global maximum of this function is found to be located at $y=1$ and
$x=\cos\theta = (\cos\frac{6\pi}{7}- \cos\frac{4\pi}{7})/(1+
|\cos\frac{4\pi}{7}|)$ which corresponds to the {\bf Sept}. 

The next lower maximum occurs at $y=1$ i.e. $\alpha=0$ and
$x= cos\theta = (\cos\frac{2\pi}{7}- \cos\frac{6\pi}{7})/(1+
|\cos\frac{6\pi}{7}|)$ which corresponds to the UPB
of DiVincenzo et al\cite{diVincenzo1} with $m=3$.
Note that, in this case, there is a third local maximum at $x=0.469$.
Thus we find that, within this family, the ${\bf Sept}$ is the strongest
UPB.

To conclude, we have introduced a rather natural notion of strength of a
a UPB and suggested that it could, in turn, provide an indirect but
calculable measure of the entropy of bound entanglement of formation
of the associated bound entangled state and perhaps also of the degree of
distinguishability of the members of a UPB under local operations and
classical communication (if such a notion could be quanified). This
measure, besides bringing out the privilleged role of the {\bf Pyramid}
and {\bf Sept} in their respective families, has two desirable properties:
\begin{itemize}
\item It vanishes when any of the parameters associated with a family of
UPB's takes a value such that the corresponding set of vectors ceases to be
a UPB.
\item The strength of a UPB obtained by taking tensor product of two
UPB's is equal to the product of the individual strengths. 
\end{itemize}
Finally, we note that an optical realization of unextendible product bases
has already been proposed \cite{zeilinger} and it will perhaps not be too
long before it is experimentally implemented. We  hope that the
notion of strength of a UPB introduced in this work will  be useful in this
context and will initiate further activity in this direction.

\end{document}